\documentclass[a4paper,11pt]{article}
\usepackage{graphicx}
\usepackage{fullpage}
\pdfoutput=1 

\usepackage{jheppub} 

\usepackage{booktabs}
\usepackage{subcaption}
\usepackage{framed}
\usepackage{multirow}
\usepackage{array}
\usepackage{lscape}
\usepackage{rotating}
\usepackage{float}

\let\jnfont=\rm
\def\NPB#1,{{\jnfont Nucl.\ Phys.\ B }{\bf #1},}
\def\PLB#1,{{\jnfont Phys.\ Lett.\ B }{\bf #1},}
\def\EPJC#1,{{\jnfont Eur.\ Phys.\ Jour.\ C }{\bf #1},}
\def\PRD#1,{{\jnfont Phys.\ Rev.\ D }{\bf #1},}
\def\PRL#1,{{\jnfont Phys.\ Rev.\ Lett.\ }{\bf #1},}
\def\MPLA#1,{{\jnfont Mod.\ Phys.\ Lett.\ A }{\bf #1},}
\def\JPG#1,{{\jnfont J.\ Phys.\ G}{\bf #1},}
\def\CTP#1,{{\jnfont Commun.\ Theor.\ Phys.\ }{\bf #1},}
\def\ZPC#1,{{\jnfont Z.\ Phys.\ C }{\bf #1},}
\def\JHEP#1,{{\jnfont JHEP \ }{\bf #1},}
\def\Rv{\not{\hbox{\kern-1pt $R$}}}
\def\p{\not{\hbox{\kern-3pt $p$}}}

\newcommand{\lsim}{\mathrel{\lower4pt\hbox{$\sim$}}\hskip-10.5pt\raise1.6pt\hbox{$<$}\;}
\newcommand{\gsim}{\mathrel{\lower4pt\hbox{$\sim$}}\hskip-10.5pt\raise1.6pt\hbox{$>$}\;}

\newcommand{\beq}{\begin{eqnarray}}
\newcommand{\eeq}{\end{eqnarray}}
\newcommand{\bpmatrix}{\begin{pmatrix}}
\newcommand{\epmatrix}{\end{pmatrix}}

\newcommand{\ba}{\begin{array}}
\newcommand{\ea}{\end{array}}

\newcommand{\be}{\begin{equation}}
\newcommand{\ee}{\end{equation}}

\title{\sffamily Explanation of the ATLAS Z-peaked excess in the NMSSM}

\author[a,b]{Junjie Cao,}
\author[a]{Liangliang Shang,}
\author[c]{Jin Min Yang,}
\author[c]{Yang Zhang}

\affiliation[a]{Department of Physics,
                Henan Normal University, Xinxiang 453007, China}
\affiliation[b]{Department of Applied Physics, Xi'an Jiaotong University, Xi'an 710049, China}
\affiliation[c]{State Key Laboratory of Theoretical Physics, Institute of Theoretical Physics,
                Academia Sinica, Beijing 100190, China}

\emailAdd{junjiec@itp.ac.cn}
\emailAdd{shlwell1988@gmail.com}
\emailAdd{jmyang@itp.ac.cn}
\emailAdd{zhangyang@itp.ac.cn}

\abstract{
Recently the ATLAS collaboration reported a $3\sigma$ excess in the leptonic-$Z+jets+E_{T}^{miss}$ channel.
This may be interpreted in the Next-to-Minimal Supersymmetric Standard Model (NMSSM) by gluino pair production
with the decay chain $\tilde{g} \to q \bar{q} \tilde{\chi}_2^0 \to q \bar{q} Z \tilde{\chi}_1^0$,
where $\tilde{\chi}_1^0$ and $\tilde{\chi}_2^0$ denote
the lightest and the next-to-lightest neutralinos with singlino and bino as their dominant components respectively.
After exploring the relevant parameter space of the NMSSM by considering the constraints
from the ATLAS searches for $jets + E_{T}^{miss}$ signals, we conclude that
the NMSSM is able to explain the excess at $1 \sigma$ level with the number of the signal
events reaching its measured central value in optimal cases, and the best explanation comes from
a compressed spectrum such as $m_{\tilde{g}} \simeq 650 {\rm GeV}$, $m_{\tilde{\chi}_2^0} \simeq 565 {\rm GeV}$ and
$m_{\tilde{\chi}_1^0} \simeq 465 {\rm GeV}$. We also check the consistency of the ATLAS results with the null result of the
CMS on-$Z$ search. We find that under the CMS limits at $95\%$ C.L., the event number of the ATLAS  on-$Z$ signal
can still reach 11 in our scenario, which is about $1.2 \sigma$ away from the measured central value.
}

\begin{document}
\maketitle  \indent
\newpage

\section{Introduction}
Since the discovery of a Higgs-like particle in 2012 \cite{Higgs-discovery}, the search for new physics beyond the Standard Model (SM) has been the most important task of the LHC. So far this search has covered a wide range of possible signatures
of new physics, especially the signals of supersymmetry (SUSY), which are characterized by various combinations of
missing transverse energy $E_{T}^{miss}$ plus jets and/or leptons (electrons or muons).  In this direction, one impressive
result is that the ATLAS collaboration recently reported a $3\sigma$
excess in the channel of two leptons with an invariant mass located around $m_Z$, large $E_T^{miss}$ and at least two
jets \cite{ATLAS-Z-Excess}.
For $20.3 fb^{-1}$ of integrated luminosity at a center-of-mass energy of $8 {\rm TeV}$, the number of the observed
leptonic-$Z$ events is 29 (summing over electron and muon pairs) in comparison with $10.6 \pm 3.2$ events expected
in the SM \cite{ATLAS-Z-Excess}.

During past weeks, several works appeared to interpret the excess in SUSY \cite{Explanation-1,Explanation-2,Explanation-3,Explanation-4,Explanation-5} and all of them employed gluino pair production process. As pointed out in \cite{Explanation-1},
in order to enhance the signal rate to meet the ATLAS data, the gluino must be relatively light,
$m_{\tilde{g}} \lesssim 1.2 {\rm TeV}$, and meanwhile have a rich $Z$-boson yield in its decay.
The latter requirement can be satisfied in the General Gauge Mediation model (GGM) with a very light gravitino
$\tilde{G}$ as the lightest supersymmetric particle (LSP) \cite{GGM}.
In this framework, due to the very weak couplings of $\tilde{G}$ to matter fields, gluino prefers to decay firstly
into a neutralino, and the subsequent decay of the neutralino into $\tilde{G}$ is able to produce one or more
$Z$-bosons \cite{Explanation-1}.
The simplest decay chain can be written as $\tilde{g} \to q \bar{q} \tilde{\chi}_1^0 \to
q \bar{q} Z \tilde{G}$, where $\tilde{\chi}_1^0$ denotes the lightest neutralino and the first step decay
proceeds via exchanging an off-shell squark. It should be noted that, since $m_{\tilde{g}} \gg m_Z$, one has
following mass relations: $m_{\tilde{g}} \gg m_{\tilde{\chi}_1^0} \gtrsim m_Z$ or $m_{\tilde{g}} \gtrsim m_{\tilde{\chi}_1^0} \gg m_Z$.
In either case, the gluino cascade decay will produce at least two hard jets, which are from the first step decay
$\tilde{g} \to q \bar{q} \tilde{\chi}_1^0$ or from the hadronic decay of an energetic $Z$ boson. As a result, this
explanation could be tightly constrained by the search for SUSY in events with jets plus $E_{T}^{miss}$.
Moreover, as illustrated in \cite{Explanation-2}, this explanation can not reproduce
the shape of the $E_T^{miss}$ distribution measured by the ATLAS collaboration.

Given the deficiencies of the GGM in explaining the excess, we in this work consider an alternative simple scenario
realized in the Next-to Minimal Supersymmetric Standard Model (NMSSM) \cite{NMSSM}. Our scenario involves a
singlino-dominated LSP $\tilde{\chi}_1^0$ and a bino-dominated next-to-lightest supersymmetric particle (NLSP)
$\tilde{\chi}_2^0$, and the decay chain $\tilde{g} \to q \bar{q} \tilde{\chi}_2^0 \to q \bar{q} Z
\tilde{\chi}_1^0$. Similar to the GGM explanation, the LSP  in our scenario has very weak couplings to matter fields,
but different from a very light $\tilde{G}$ in GGM (which is required by the prompt decay of the NLSP \cite{GGM-Light-G}),
now the LSP may be massive.
This feature enables us to choose a slightly compressed SUSY spectrum
$m_{\tilde{g}} \sim m_{\tilde{\chi}_2^0} \sim m_{\tilde{\chi}_1^0} + m_Z$
to evade the constraints from the SUSY searches. Consequently, gluino as light as $700 {\rm GeV}$ is still
experimentally allowed, which is helpful to explain the excess (see discussion below). Another advantage of our
scenario is that it can reproduce the $E_T^{miss}$ and $H_T$ distributions measured by the ATLAS collaboration.
This has been recently emphasized in \cite{Explanation-2}.

About our scenario, two points should be noted. One is that, since $Z$ boson decays dominantly hadronically,  in most
cases the gluino cascade will produce multiple jets. Considering that the ATLAS collaboration has performed several
comprehensive searches for the jets + $E_T^{miss}$ signals, and each of the searches focused on multiple signal regions
which are sensitive to different configurations of SUSY spectrum \cite{ATLAS-6jet-old,ATLAS-6jet-new,ATLAS-7jet},
one should consider these searches by reproducing them through detailed simulations and then using the corresponding
experimental data as input to limit SUSY.
Studying the capability of our scenario under these constraints to
interpret the excess is the main aim of this work.
The other is that the CMS collaboration recently also reported its analysis on the
leptonic-$Z+jets+E_T^{miss}$ signal \cite{CMS-Leptonic-Z}, which should have the same physical origin
as the ATLAS excess in our scenario. Since the CMS observed no excess, one should check the consistency
of the two searches by noting that they are based on different cuts and different detectors.
This is another aim of this work.

This work is organized as follows. In Section II, we introduce our scenario and illustrate
its key features. Then in Section III, we describe the strategies of the ATLAS and CMS searches
for the leptonic-$Z+jets+E_T^{miss}$ signal as well as other searches for $jets+E_T^{miss}$ signals.
In Section IV, we first show the dependence of various signals on SUSY parameters in our scenario.
Then we scan the relevant SUSY parameters by considering the constraints from the SUSY searches,
and in the surviving parameter space, we investigate the capability of our scenario in explaining
the ATLAS Z-peaked excess and check the consistency of the measurements from the two collaborations
on the leptonic-$Z+jets+E_T^{miss}$ signal.  Finally, we draw our conclusion in Section V.
The validations of our simulations on the ATLAS searches for the leptonic-$Z+jets+E_T^{miss}$ signal
are presented in the appendix.

\section{A potential explanation of the excess in the NMSSM}
The NMSSM is the simplest singlet extension of the Minimal Supersymmetric Standard Model (MSSM)
and its superpotential is given by \cite{NMSSM}
\begin{eqnarray}
W_{NMSSM} = W_{MSSM} + \lambda \hat{S} \hat{H_u} \cdot \hat{H_d} + \xi_F \hat{S} + \frac{1}{2} \mu^\prime \hat{S}^2  + \frac{\kappa}{3} \hat{S}^3,
\end{eqnarray}
where $W_{MSSM}$ is the superpotential of the MSSM, $\hat{H_u}$ and $\hat{H_d}$ are MSSM Higgs superfields, $\hat{S}$ is a
gauge singlet superfield, $\lambda$ and $\kappa$ are dimensionless Yukawa couplings, $\mu^\prime$ is a supersymmetric mass
and $\xi_F$ with dimension of squared mass parameterizes a tadpole term.

Due to the addition of the singlet field,  the NMSSM predicts three CP-even Higgs
bosons, two CP-odd Higgs bosons and five neutralinos. The mass matrix of the neutralinos in the
bases $(-i \tilde{B}^0, - i \tilde{W}^0,
\tilde{H}_u^0, \tilde{H}_d^0, \tilde{S}^0)$ is given by\cite{NMSSM}
\begin{eqnarray}
\cal{M} = \left( \begin{array}{ccccc}
  M_1 & 0 &  \frac{e v_u}{\sqrt{2} c_w} & -\frac{e v_d}{\sqrt{2} c_w} & 0  \\
  0& M_2 & -\frac{e v_u}{\sqrt{2} s_w} & \frac{e v_d}{\sqrt{2} s_w} & 0  \\
   \frac{e v_u}{\sqrt{2} c_w} & -\frac{e v_u}{\sqrt{2} s_w} & 0 & -\mu_{eff} & -\lambda v_d  \\
    -\frac{e v_d}{\sqrt{2} c_w} & \frac{e v_d}{\sqrt{2} s_w} &  -\mu_{eff} & 0& -\lambda v_u \\
    0 & 0 &  -\lambda v_d & -\lambda v_u & 2 \kappa s + \mu^\prime
   \end{array} \right),
\end{eqnarray}
where $c_w = \cos \theta_W$, $M_1$ and $M_2$ are soft gaugino masses, $v_u= v\sin \beta$ and $v_d = v \cos \beta$ are vacuum expectation values (vev)
of the fields $H_u$ and $H_d$ respectively, and $\mu_{eff} = \mu + \lambda s $ with $s$ denoting the vev of
the singlet scalar field $S$. This matrix indicates that the mixings between the bases depend on the parameters $\tan \beta$ and
$\lambda$, and for a singlino-dominated LSP,  its gaugino components should be very small, while its Higgsino components may be sizeable for a large $\lambda$.

The interactions of the neutralinos take following form \cite{NMSSM}
\begin{eqnarray}
{\cal{L}}_{\tilde{\chi}^0} & = & \tilde{u}_L^\ast \bar{{\tilde \chi}}_j^0 \Big[ \frac{-e}{\sqrt{2}s_w c_w}( {1 \over 3} N_{1j}s_w
+ N_{2j}c_w )P_L -y_uN_{4j}^\ast P_R \Big] u
\nonumber \\
&& + \  \tilde{d}_L^\ast \bar{\tilde{\chi}}_j^0 \Big[ {{-e} \over {\sqrt{2}s_w c_w}}( {1 \over 3} N_{1j}s_w
- N_{2j}c_w )P_L +y_dN_{3j}^\ast P_R \Big]d  \nonumber \\
&& + \  \tilde{u}_R^\ast {\bar{\tilde{\chi}}}_j^0 \Big[ {{2\sqrt{2}e} \over {3 c_w}} N_{1j}^\ast P_R -y_uN_{4j} P_L \Big] u +
\tilde{d}_R^\ast {\bar{\tilde{\chi}}}_j^0 \Big[ {{-\sqrt{2}e} \over {3 c_w}} N_{1j}^\ast P_R +y_dN_{3j} P_L \Big]d  \nonumber  \\
&& +  \  {e \over {s_w c_w}}Z_\mu \bar{\tilde{\chi}}_i^0 \gamma^\mu({\cal{O}}_{ij}^L P_L + {\cal{O}}_{ij}^R P_R){\tilde{\chi}}_j^0 +  h_u \bar{\tilde{\chi}}_i^0
({\lambda \over \sqrt{2}} \Pi^{45}_{ij} - {g_1 \over 2} \Pi^{13}_{ij} + {g_2 \over 2} \Pi^{23}_{ij} ){\tilde{\chi}}_j^0 + \cdots
\end{eqnarray}
where $N$ is the rotation matrix to diagonalize the neutralino mass matrix with its element denoted by
$N_{ij}$ ($i,j=1,\cdots, 5$),
${\cal{O}}_{ij}^L= - {\cal{O}}_{ij}^{R \ast} = {-}{1 \over 2}N_{i3}N_{j3}^\ast + {1 \over 2}N_{i4}N_{j4}^\ast $, and
$\Pi^{ab}_{ij}= N_{ia}N_{jb}+N_{ib}N_{ja}$. This Lagrangian indicates that the interactions of the neutralinos with
a light quark and those with a Z-boson are determined by their gaugino components and Higgsino components respectively,
while their interactions with a Higgs boson are decided by the both and there may exist cancelation between the two
contributions. These features are helpful in understanding our scenario.

As mentioned in last section, the key issues in explaining the leptonic-$Z$ excess are how to improve the $Z$ yield
rate in gluino decay and how to escape the constraints from the ATLAS direct searches for SUSY. Our scheme is as follows:
\begin{itemize}
\item We set the LSP to be the singlino-dominated neutralino, the NLSP to be the bino-dominated neutralino, and $M_2, \mu_{eff} > m_{\tilde{g}}$. This setup
guarantees that the rate of the gluino decay into the $\tilde{\chi}_2^0$ (via an intermediate off-shell squark) is much larger than that of the decay into
the $\tilde{\chi}_1^0$. Meanwhile, since the NLSP is not accompanied by a light chargino, the gluino may decay into the $\tilde{\chi}_2^0$ at a
rate of approximate $100\%$.
\item We assume $m_{\tilde{t}}, m_{\tilde{b}} \gg m_{\tilde{q}} > m_{\tilde{g}}$ where $q$ denotes light quarks. This assumption ensures that the gluino
decays dominantly into $q\bar{q} \tilde{\chi}_2^0$ state at first step. We suppress the decay chain $\tilde{g} \to t\bar{t} \tilde{\chi}_2^0 \to t\bar{t} Z \tilde{\chi}_1^0$ due to its very low efficiency in getting the desired signal (We obtain this conclusion by detailed simulations).
We assume a relatively heavy $\tilde{q}$ since it is favored by current LHC search for SUSY.

As will be shown later, we are particularly interested in the situation $m_{\tilde{g}} \sim  m_{\tilde{\chi}_2^0}$. In this case,
the assumption on the mass spectrum  may be relaxed to be
$m_{\tilde{t}}, m_{\tilde{b}}, m_{\tilde{q}} > m_{\tilde{g}}$ and $m_{\tilde{g}} < m_{\tilde{\chi}_2^0} + 2 m_t$.

\item  We require that the decay $ \tilde{\chi}_2^0 \to Z \tilde{\chi}_1^0$ proceeds at a rate of approximate $100\%$.
This can be realized by setting $ m_Z < m_{\tilde{\chi}_2^0} - m_{\tilde{\chi}_1^0}  \leq m_h $ or by choosing specific SUSY
parameters such that the coupling $h \bar{\tilde{\chi}}_1^0 \tilde{\chi}_2^0$ is suppressed. With the package NMSSMTools\cite{NMSSMTools}, we
numerically checked  that, by tuning the parameters of the NMSSM, the rate of  $ \tilde{\chi}_2^0 \to Z \tilde{\chi}_1^0$
can be significantly larger than that of $ \tilde{\chi}_2^0 \to h \tilde{\chi}_1^0$ even when
$m_{\tilde{\chi}_2^0} - m_{\tilde{\chi}_1^0}  > m_h $.

\item We employ a moderately compressed spectrum $m_{\tilde{g}} \sim m_{\tilde{\chi}_2^0} \sim m_{\tilde{\chi}_1^0} + m_Z$ to avoid the constraints from the
LHC searches for SUSY.
\end{itemize}

About our scenario, we have more explanations. One is that the measured SM-like Higgs boson mass in our scenario can
be satisfied by choosing appropriate values of $\lambda$, $\tan \beta$ and also third generation squark
masses \cite{Higgs-Mass-1,Higgs-Mass-2,Higgs-Mass-3,Higgs-Mass-4}
and the correct dark matter relic density can be achieved by the annihilate channels
$\tilde{\chi}_1^0  \tilde{\chi}_1^0 \to H_i/A_j \to X Y$ with $X Y$ being any possible final states \cite{NMSSM-Dark-Matter}.
The other is that, in our scenario, we ad hoc forbid the possibility of $M_2, \mu_{eff} < m_{\tilde{g}}$. The main reason
is that in that case, the gluino decay chain may become more complicated, and channels such as
$\tilde{g} \to q \bar{q}^\prime \tilde{\chi}_i^\pm \to q \bar{q}^\prime W^\pm \tilde{\chi}_j^0 \to q \bar{q}^\prime W^\pm Z \tilde{\chi}_1^0$ are open. Then
one has to scrutinize the vast parameter space of the NMSSM to find parameter points where the $Z$ yield rate is high. Since this process involves the
simulation of the gluino pair production at the LHC, it is very time-consuming in calculation by cluster because the decay chains are lengthy.
So our scenario should be regarded as a simplest attempt to explain the excess,
which only involves the parameters $m_{\tilde{g}}$, $m_{\tilde{\chi}_1^0}$ and  $m_{\tilde{\chi}_2^0}$.

\section{Simulations and cuts}
In order to discuss the constraints of the LHC searches for SUSY on our scenario and also calculate the event number of the leptonic-$Z$
$+ E_T^{miss}$ signal, we implement the corresponding experimental searches in the package CheckMATE-1.2.0 \cite{checkmate}, which have
tuned the package Delphes3.0.10 \cite{delphes} to simulate the behavior of the ATLAS (CMS) detector. The validations of our simulations
are presented in Appendix for the benchmark points provided by the experimental groups, and we find that our results
agree with the corresponding experimental analyses at $20\%$ level. Throughout this work, the events in our
simulation are generated by MG5\_aMC\cite{mg5} which includes Pythia\cite{pythia} for parton showering and hadronization, and
the package Prospino \cite{prospino} is used to calculate the NLO cross sections of the gluino pair production.

In the following, we only briefly describe the cut flows of these experiments. Detailed information about them can be found from the corresponding
experimental reports.

\subsection{ATLAS search for leptonic-$Z+jets+E_T^{miss}$ signal}

This search concentrated on the events with a same flavor opposite-sign (SFOS) dilepton pair, jets and $E_T^{miss}$ at 8-TeV LHC \cite{ATLAS-Z-Excess}.
One intriguing result of this search comes from the signal region SR-Z where the invariant mass of the lepton pair locates around $m_Z$. Explicitly
speaking,  in contrast with $4.2 \pm 1.6$ electron pair events and $6.4 \pm 2.2$ muon pair events expected for the SM background,
16 events and 13 events are observed respectively in this region. This implies a $3 \sigma$ excess of the observed events over
the background, and may be regarded
as a hint of SUSY.  So in the following we try to explain this excess with our scenario. The signal region SR-Z is defined by

\begin{itemize}
  \item The first two leading leptons are SFOS, their transverse momentums are larger than $25 {\rm GeV}$ and $10 {\rm GeV}$ respectively,
  and meanwhile their invariant mass satisfies $81GeV < m_{ll} < 101GeV$.
  \item More than one jets have $P_T>35GeV$ and $|\eta|<2.5$.
  \item $E_T^{miss}>225GeV$ and $H_T>600 {\rm GeV}$, where $H_T$ is defined as the scalar sum of the transverse momenta for
  all signal jets and the two leading leptons.
  \item The azimuthal opening angle between each of the leading two jets and $E_{T}^{miss}$ should be larger than 0.4, i.e. $\Delta\phi(jet_{1,2},E_T^{miss})>0.4$.
\end{itemize}

\subsection{ATLAS search for  $> 6~jets + E_T^{miss}$ signal}

\begin{table}[t]\caption{Definition of signal regions in ATLAS search on  $> 6~jets + E_T^{miss}$ signal \cite{ATLAS-7jet}. Here the jet, the pseudo-rapidity
$\eta$ and multiplicity all  refer to the $R=0.4$ jets. Composite jets with a larger radius parameter $R=1.0$ are used in the
multi-jet + $M^{\sum}_J$ stream when constructing the quantity $M^{\sum}_J$ (see text). \label{table1}}\centering
\begin{tabular}{|l|c|c|c|c|c|c|c|c|}
\hline
                         & \multicolumn{5}{c|}{Multi-jet + flavour stream}                 & \multicolumn{3}{c|}{Multi-jet + $M^{\sum}_J$ stream} \\ \hline
Name                     & 8j50        & 9j50        & $\geq$10j50 & 7j80        & $\geq$8j80  & $\geq$8j50      & $\geq$9j50      & $\geq$10j50\\ \hline
Jet $P_T^{min}${[}GeV{]} & \multicolumn{3}{c|}{50}               & \multicolumn{2}{c|}{80}     & \multicolumn{3}{c|}{50}                              \\ \hline
Jet count                & $=$8        & $=$9        & $\geq$10    & $=$7        & $\geq$8     & $\geq$8         & $\geq$9         & $\geq$10         \\ \hline
b-jet count              & 0,1,$\geq$2 & 0,1,$\geq$2 &             & 0,1,$\geq$2 & 0,1,$\geq$2 & \multicolumn{3}{c|}{}                                \\ \hline
$M^{\sum}_J${[}GeV{]}    & \multicolumn{5}{c|}{}                                           & \multicolumn{3}{c|}{$>340,>420$}                     \\ \hline
\end{tabular}
\end{table}

To search for final states with large jet multiplicities and $E_T^{miss}$, ATLAS used two independent streams, i.e. multi-jet + flavor stream and multi-jets+$M^{\sum}_J$ stream, in its anlaysis \cite{ATLAS-7jet}.

In the multi-jet + flavour stream, the number of jets with $|\eta|<2$ and $p_T$ above a threshold ($50 {\rm GeV}$ or $80 {\rm GeV}$)
is firstly determined for a given event. Then events are categorized by the number of jets, and further subdivided by the number of
jets that are $b$-tagged. After such treatments, totally thirteen signal regions are defined, which are given in
Table.\ref{table1}.

Analysis of the multi-jet + $M^{\sum}_J$  stream seems more complicated. As the first step, the number of ($R=0.4$) jets with $|\eta|<2.8$ and
$p_T$ above 50\,GeV is determined for a given event, and like the former stream, events are classified by their jet numbers.
As the second step, the four-momenta of the $R=0.4$ jets are used as inputs for a second iteration of the anti-$k_t$ jet algorithm \cite{antikt},
which now adopts a larger jet radius parameter, $R= 1.0$. The resulting larger objects are called as composite jets. Then a selection
variable $M^{\sum}_J$  is defined by
\begin{equation}
M^{\sum}_J \equiv \sum_{j} m_{j}^{R=1.0},
\end{equation}
where the sum is over the mass of the composite jets, and two signal regions are defined by the threshold of $M^{\sum}_J$
($340 {\rm GeV}$ or $420 {\rm GeV}$). In this stream, there are
totally six signal regions (see Table.\ref{table1}).

We mention by the way that both the streams veto events containing any isolated electron or muon candidates with $P_T > 10GeV$ at the start of the analysis,
and require $E_T^{miss}/\sqrt{H_T} > 4GeV^{1/2}$ at the end of the cut flows, where $H_T$ is the scalar sum of the transverse momenta of all jets with $P_T > 40GeV$.

\subsection{ATLAS search for  $2\sim6~jets + E_T^{miss}$ signal}

The search for events with $2\sim6~jets + E_T^{miss}$ at $8$-TeV LHC with a total integrated luminosity of $20.3fb^{-1}$ has been carried out
in \cite{ATLAS-6jet-old} and also in \cite{ATLAS-6jet-new}. Compared with the preliminary search described in \cite{ATLAS-6jet-old}, the updated search
in \cite{ATLAS-6jet-new} defined more signal regions to optimize the search. For example, it has been shown that the signal regions  $2jW$ and $4jW$
are able to improve greatly the sensitivity of the ATLAS search to models predicting the cascade $\tilde{g} \to q \bar{q}^\prime \tilde{\chi}_1^{\pm} \to q \bar{q}^\prime W \tilde{\chi}_1^0$. The main cuts of this search are:
\begin{itemize}
  \item $E_T^{miss}>160 {\rm GeV}$.
  \item $P_T(j_1)>130GeV$ for the leading jet, and $P_T(j)>60GeV $ for the other jets. An exception is the signal region
  $4 jW $ where $ P_T(j)>40GeV $ is required.
  \item The azimuthal angles between jets and the $E_{T}^{miss}$ satisfy $\delta \phi(jet_{1,2,(3)},E_T^{miss})_{min}>0.4$ and
  $\delta \phi(jet_{i>3},E_T^{miss})_{min}>0.2$.
  \item  A variety of signal regions is then defined by the number of jets,  the values of $E_T^{miss}/\sqrt{H_T}$, $E_T^{miss}/m_{eff}(N_j)$ and
  $m_{eff}(incl)$. Here $H_T$, $m_{eff}(N_j)$ and $m_{eff}(incl.)$ are defined as the scalar sum of the transverse momenta for all
  $P_T > 40GeV$ jets, the leading $N_j$ jets and $E_T^{miss}$, and all jets with $P_T > 40 {\rm GeV}$ and  $E_T^{miss}$ respectively.
\end{itemize}

 In practice, we also consider the search in \cite{ATLAS-6jet-old}. This search has been implemented in the CheckMATE
 by the package authors, and we have verified its correctness. Since this search is significantly weaker
 than that of \cite{ATLAS-6jet-new} in constraining SUSY parameter space, we do not list its cuts here.

\subsection{CMS search for leptonic-$Z$ $+jets + E_T^{miss}$ signal}

\begin{table}[t]\caption{Six bins of the CMS dedicated counting experiment for events with an on-shell Z boson \cite{CMS-Leptonic-Z}.
For each bin, the 95\% CL upper limits on the number of signal events, $S_{obs}^{95}$, are also presented.\label{tab:CMS95}}\centering
\begin{tabular}{|l|r|r|r|r|r|r|}
\hline
$N_{jets}$        & \multicolumn{3}{c|}{$\geq2$}                                  & \multicolumn{3}{c|}{$\geq3$}                 \\ \hline
$E_T^{miss}[GeV]$ & 100-200      & 200-300       & \textgreater300 & 100-200    & 200-300      & \textgreater300 \\\hline
$N_{data}$        & 1187         & 65            & 7               & 490        & 35           & 6               \\\hline
$N_{bkg}$         & 1204$\pm$106 & 74.5$\pm$11.3 & 12.8$\pm$4.3    & 478$\pm$43 & 39.2$\pm$6.6 & 5.3$\pm$2.3     \\\hline
$S_{obs}^{95}$    & 207          & 20            & 7.6             & 89         & 16.1         & 8               \\ \hline
\end{tabular}
\end{table}

In \cite{CMS-Leptonic-Z}, the CMS collaboration carried out a dedicated on-Z counting experiment to search for leptonic-$Z$
$+jets + E_T^{miss}$ signal. This signal is same as that of the ATLAS search in \cite{ATLAS-Z-Excess}, but different from the ATLAS result,
the CMS collaboration saw no excess. Since the two signals have same physical origin in our scenario, they should be correlated.
So in our work, we consider the CMS search as a possible constraint on our interpretation of the excess. The main strategies of the CMS search
are given by
\begin{itemize}
\item There exists at least one SFOS dilepton pair (electron or muon pair) with $P_T>20GeV$ and $|\eta|<2.4$ for each lepton,
and the invariant mass of the lepton pair satisfies  $81<m_{ll}<101GeV$.
\item Six bins are defined by the value of $E_T^{miss}$ and the number of jets with $P_T>40GeV$ and $|\eta|<3.0$.
Here we present some information of these bins in TABLE \ref{tab:CMS95}.
\end{itemize}

About the CMS search, we note that, unlike above ATLAS searches, it does not provide the $95\%$ C.L. upper limits on the number
of signal events for each bin, which was denoted as $S_{obs}^{95}$ in literature \cite{checkmate}. In order to discuss the constraint
of such a search, we calculate these limits by the asymptotic $CL_s$ prescription \cite{CLs}, and present the corresponding
results in TABLE \ref{tab:CMS95}. As we will show below, the bins with $N_j \geq 2,3 $ and $ 200 < E_T^{miss} < 300 $ usually
put the strongest limitation on any explanation of the excess, which is different from previous analysis \cite{Explanation-2}.

\section{Numerical results and discussions}

\begin{figure}[t]
  \centering
  \includegraphics[width=15cm]{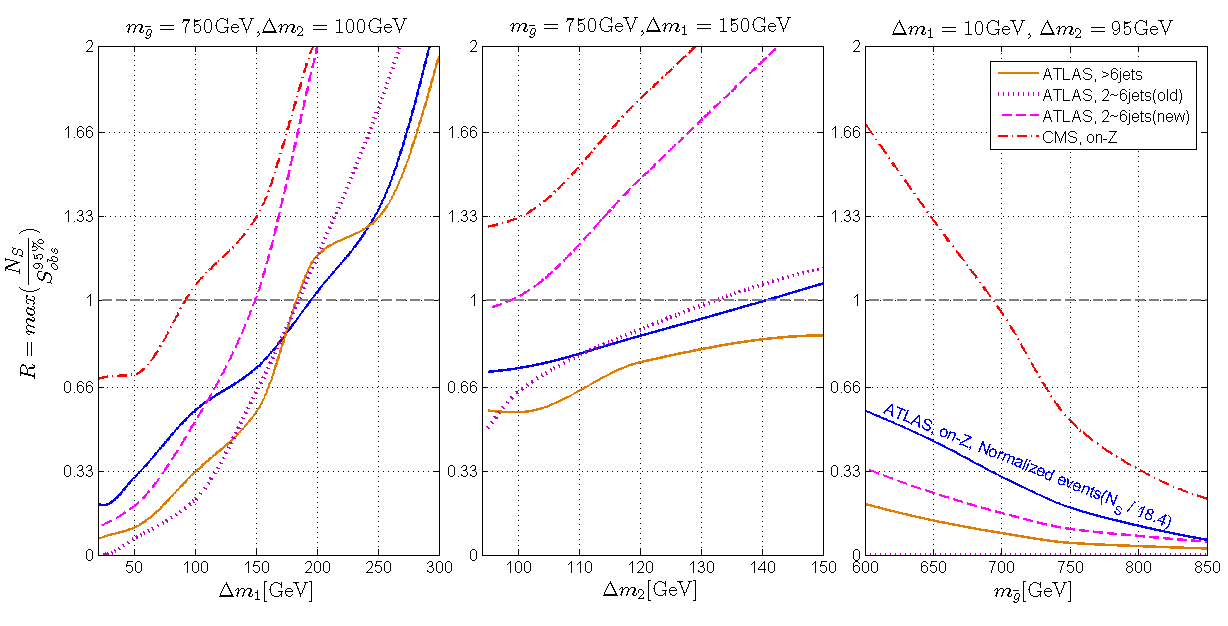}\\
  \caption{The dependence of $R$s for different searches on $\Delta m_1$, $\Delta m_2$ and $m_{\tilde{g}}$.
  Here $\Delta m_1 = m_{\tilde{g}} - m_{\tilde{\chi}_2^0} $ denotes the mass splitting between $\tilde{g}$ and
  $\tilde{\chi}_2^0$ and $\Delta m_2 = m_{\tilde{\chi}_2^0} - m_{\tilde{\chi}_1^0}$ is that
  between $\tilde{\chi}_2^0$ and $\tilde{\chi}_1^0$. For the ATLAS leptonic-$Z$ analysis, $R$ is the normalized lepton pair
  event number, $R = N_{ll}/18.4$ with $18.4$ representing the central value of the excess, while for any of the other
  searches, $R = max(N_{S,i}/  S_{obs,i}^{95\%})$ where $N_{S,i}$ is the event number for $i$ signal region in the
  search, $S_{obs,i}^{95\%}$ is its $95\%$ upper limit, and the $max$ is over all signal regions in the search.
  In the latter case, $R<1$ means that the parameter point is allowed by the corresponding search at $95\%$ C.L..}\label{fig1}
\end{figure}

As we introduced in section II, the branching ratio of $\tilde{g} \to q \bar{q} \tilde{\chi}_2^0 \to q \bar{q} Z \tilde{\chi}_1^0$ in our scenario can approach
$100 \%$ by setting the SUSY parameters other than the masses $m_{\tilde{g}}$, $m_{\tilde{\chi}_2^0}$ and $ m_{\tilde{\chi}_1^0}$ at certain values.
In this case, the event number of the gluino pair production with a certain final state is solely determined by the masses, or equivalently by $m_{\tilde{g}}$,
$\Delta m_1 = m_{\tilde{g}} - m_{\tilde{\chi}_2^0} $ and $\Delta m_2 = m_{\tilde{\chi}_2^0} - m_{\tilde{\chi}_1^0}$. In this section,
we investigate the constraints of the SUSY searches on the masses. For this end, we define for each experimental search the ratio
$R = max (N_{S,i}/ S_{obs,i}^{95\%})$, where $N_{S,i}$ is the signal event number for $i$ signal region in the search,
$S_{obs,i}^{95\%}$ is its $95\%$ upper limit, and the $max$ is over all signal regions of the search. Obviously, only in the case of $R < 1$,
the corresponding mass spectrum is allowed by the search at $95\%$ C.L.. We also investigate how well our scenario can explain the
leptonic-$Z$ excess. After considering the statistical uncertainty of the measured event number, we infer from the ATLAS search that the needed signal
event number, $N_{ll}$,  is $18.4 \pm 6.3$ (assuming the experimental data obey Poisson Distribution and adding the statistical uncertainty
and systemic uncertainty in quadrature), which corresponds to $12.1 \leq N_{ll}  \leq 24.7$ and  $5.8 \leq N_{ll}  \leq 31$ in explaining
the excess at $1 \sigma$ and $2 \sigma$ levels respectively. In following discussion, it is convenient to define the normalized leptonic-$Z$ event
number as $R_{ll} = N_{ll}/18.4$, then the $1 \sigma$ and $2 \sigma$ ranges are restated as
$0.66 < R_{ll} < 1.34 $ and $0.32 < R_{ll} < 1.68 $ respectively.

In Fig.\ref{fig1}, we show the dependence of $R$s for different searches on $\Delta m_1$, $\Delta m_2$ and $m_{\tilde{g}}$ respectively
by fixing $m_{\tilde q}=4.5$ TeV. From this figure, one can learn at least five facts.
\begin{itemize}
\item[1.] The left and middle panels indicate that, as the SUSY spectrum becomes compressed, the value of $R$ for each search decreases monotonously.
This reflects that a smaller mass splitting is able to suppress the cut efficiency in the search so that fewer SUSY signal events are finally
retained. Consequently, this case is apt to escape the constraints from the direct search for SUSY at the LHC, but meanwhile, the ATLAS leptonic-$Z$
events are suppressed too.
\item[2.] Motivated by above observation, we consider a rather compressed spectrum in the right panel, and study the dependence of $R$ on $m_{\tilde{g}}$.
The panel indicates that $R$s decrease monotonously as the gluino becomes heavy. This is because the suppression of the gluino pair production rate
with the increase of $m_{\tilde{g}}$ usually dominates over the enhancement of the cut efficiency in getting the signal events of the search, and the net
result is the dropping of the signal events.
\item[3.] All the three panels indicate that, in the case of a very compressed spectrum, the CMS search for the leptonic-$Z$ $+ jets + E_T^{miss}$
signal puts the severest constraint on SUSY among the searches. Especially, as the gluino becomes light, the signal event number in the CMS search
grew much faster than those of the other searches. As a result, the CMS search may be used to set a lower bound on $m_{\tilde{g}}$ in our scenario.
\item[4.] In general, the ATLAS preliminary search for $2\sim 6$ $jets + E_T^{miss}$ signal and that for $> 6$ $jets + E_T^{miss}$ signal are
comparable in constraining the parameter space of our scenario, and the both constraints are weaker than that from the
ATLAS updated search for $2\sim 6$ $jets + E_T^{miss}$ signal.
\item[5.] Since the values of $R$s for the experimental searches have different behaviors with the variation of the mass spectrum,
it is better to scan the masses in studying the capability of our scenario to explain the excess.
\end{itemize}

\begin{figure}[H]
  \centering
  \includegraphics[width=7.2cm]{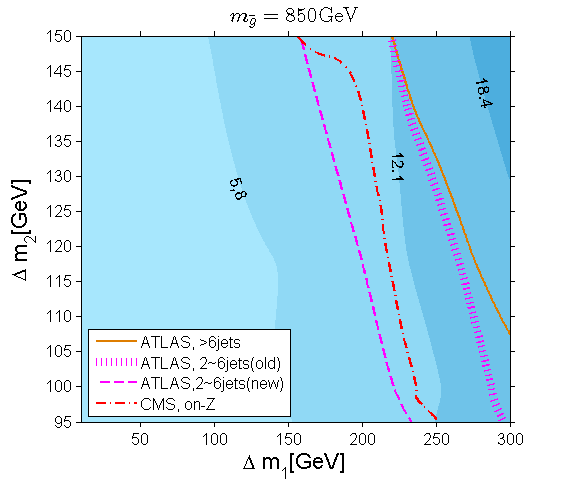} \hspace{-0.5cm} \includegraphics[width=7.2cm]{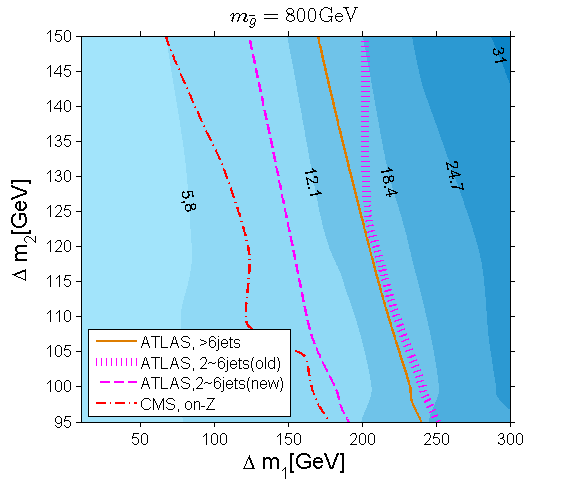}\\
  \includegraphics[width=7.2cm]{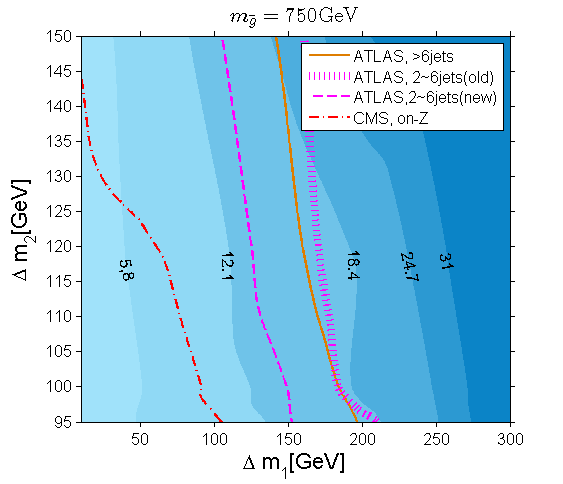} \hspace{-0.5cm} \includegraphics[width=7.2cm]{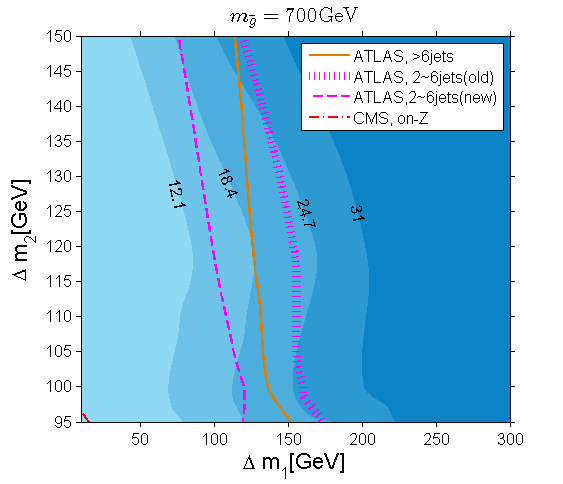}\\
   \includegraphics[width=7.2cm]{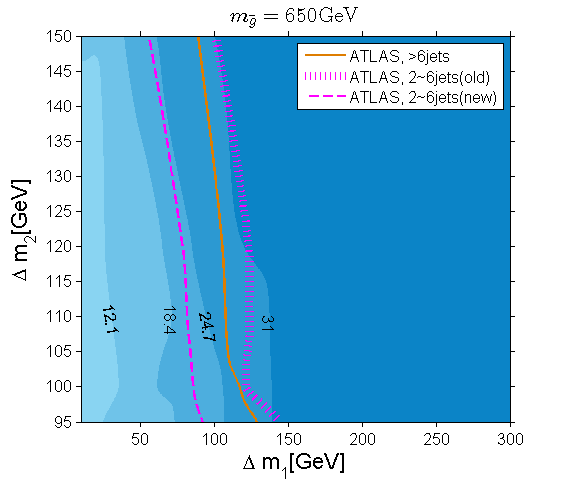} \hspace{-0.5cm} \includegraphics[width=7.2cm]{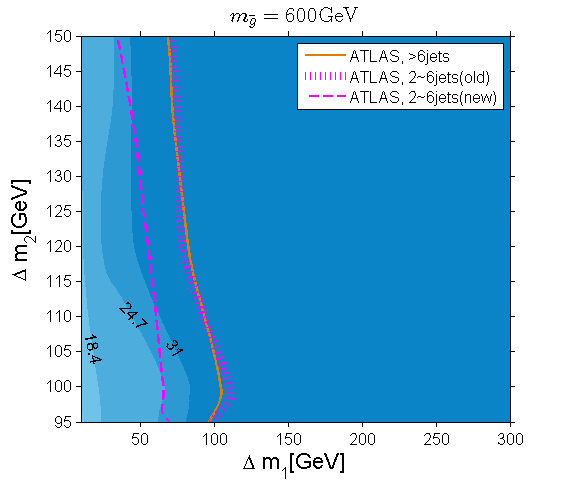}


  \caption{Constant contours of the event number for the ATLAS search for the leptonic-$Z$ $ + jets + M_T^{Miss}$ signal on
  the $\Delta m_1 - \Delta m_2$ plane for different gluino masses. Regions between line 12.1 and line 24.7 and those between
  line 5.8 and line 31 are able to explain the excess at $1\sigma $ and $2 \sigma$ levels respectively. Constraints
  from the other SUSY searches are also plotted by different types of lines  with the left side of the lines being experimental allowed.
  For the case of $m_{\tilde{g}} \lesssim 650 {\rm GeV}$, there are no regions
  surviving the constraint from the CMS search for the leptonic-$Z$ signal. }\label{fig2}
\end{figure}

In Fig.\ref{fig2}, we show the constant contours of the event number for the ATLAS search for the leptonic-$Z$ $ + jets + M_T^{Miss}$ signal on
the $\Delta m_1 - \Delta m_2$ plane for different gluino masses. Regions between line 12.1 and line 24.7 and those between line 5.8 and line 31
are able to explain the excess at $1\sigma $ and $2 \sigma$ levels respectively.  Constraints from the other SUSY searches are also plotted
by different types of lines with the left side of the lines being experimentally allowed. Solid line, dashed line and  dotted line represent the constraints
from the ATLAS searches for $>6$ $jets + E_T^{miss}$,  $2\sim6$ $jets + E_T^{miss}$ (updated) and $2\sim6$ $jets + E_T^{miss}$ (preliminary)
respectively, and they are obtained by fixing the corresponding $R$ value at 1. The CMS constraint is plotted in a similar way, but by
dash-dotted lines. In getting this figure, we scan the parameters $\Delta m_1$ and $\Delta m_2$ with a grid of
$30 {\rm GeV}$ for $\Delta m_1$ and $10 {\rm GeV}$ for $\Delta m_2$.

From Fig.\ref{fig2}, one can learn following facts:
\begin{itemize}
\item With the decrease of the gluino mass, only a more compressed spectrum can survive the constraints from the LHC searches
for SUSY. Among the constraints, the CMS search for the letponic-$Z$ $+ jets + E_T^{miss}$ becomes the most stringent constraint
for $m_{\tilde{g}} \lesssim 800 {\rm GeV}$. Especially,
in the case of $m_{\tilde{g}} = 700 {\rm GeV}$, only a very small corner on the $\Delta m_1 - \Delta m_2$ plane survives the constraint.
\item If one does not consider the CMS constraint, our scenario can explain well the ATLAS leptonic-Z excess without conflicting with the ATLAS searches
for $jets + E_T^{miss}$. The central value of the excess (18.4 events) may be achieved for $m_{\tilde{g}} \lesssim 700 {\rm GeV}$, and one of the best benchmark points is $m_{\tilde{g}} = 650 {\rm GeV}$, $m_{\tilde{\chi}_2^0} = 565 {\rm GeV}$ and $ m_{\tilde{\chi}_1^0} = 465 {\rm GeV}$.
While if one take the CMS constraint serious, we find that the leptonic-$Z$ events in the ATLAS experiment is at most 11. This tension arises since
the CMS signal and the ATLAS excess have same physical origin.
\end{itemize}

\section{Conclusion}

Recently the ATLAS collaboration reported a $3\sigma$ excess in the leptonic-$Z+jets+E_{T}^{miss}$ channel.
In this work, we intend to explain the excess by the gluino pair production process in the NMSSM. For this purpose,
we implemented this experimental analysis (and also some other useful experimental analyses)
in the package CheckMATE, and verified the reliability of our simulations.

The scenario we considered involves a singlino-like LSP, a bino-like NLSP and an approximate $100\%$ branching ratio
for the decay chain $\tilde{g} \to q \bar{q} \tilde{\chi}_2^0 \to  q \bar{q} \tilde{\chi}_1^0$. In this case, the
SUSY signal rates after cuts are essentially determined by the masses $m_{\tilde{g}}$, $m_{\tilde{\chi}_2^0}$ and
$m_{\tilde{\chi}_1^0}$.
We scanned these masses by considering the constraints from the ATLAS searches for $jets + E_{T}^{miss}$ signal, then
we concentrated on the leptonic-$Z$ signal in the experimentally allowed parameter space. We concluded that
our scenario is able to explain the excess at $1 \sigma$ level with the number of the signal events reaching its
experimental measured central value in optimal cases, and the best explanation comes from a compressed spectrum such as
$m_{\tilde{g}} \simeq 650 {\rm GeV}$, $m_{\tilde{\chi}_2^0} \simeq 565 {\rm GeV}$ and
$m_{\tilde{\chi}_1^0} \simeq 465 {\rm GeV}$. We also tested the consistency of the ATLAS result with the CMS result in
the same channel, which observed no excess on the leptonic-$Z$ production. We found that
under the CMS limits at $95\%$ C.L., the number of the ATLAS signal events can still reach 11 in our scenario,
which is about $1.2 \sigma$ away from the measured central value.

We emphasize that, although we employ a specific scenario in our work to explain the ATLAS excess, our conclusions
seem to be general. That is, the NMSSM is able to explain quite well the excess if only the constraints from the
ATLAS searches for SUSY are considered. While on the other hand, if one considers the CMS search limits,
there is a mild tension between the ATLAS excess and the CMS result.

{\bf{Note added:}} At the final stage of this work, there appeared a paper by Ulrich Ellwanger \cite{Explanation-2},
which considered a similar scenario to ours. Anyhow, our work differs from his work mainly in following three aspects:
\begin{itemize}
\item In his paper, only two benchmark points were discussed,
while we perform a scan over the relevant parameter space. Our calculation by cluster is very time-consuming.
\item We considered more constraints than in his paper, where only the ATLAS preliminary search for $2 \sim 6 jets + E_T^{miss}$
was considered. Especially, the benchmark point P2 in his paper was considered as an optimal point to explain
the excess, while in our work, we got $R=1.19$ for the updated ATLAS search for $2 \sim 6 jets + E_T^{miss}$
(with $m_{\tilde{q}}=4.5TeV$). Considering that we have about 20\% uncertainty in simulating thesearch, we infer that
the point is at the boundary of being excluded.
\item When we consider the CMS analysis, we found that the most stringent constraint comes from the bins of
$N_{jets} \gtrsim 2, 3 $ and
$ 200 {\rm GeV} < E_T^{miss} < 300 {\rm GeV}$ for most cases, instead of the bin $N_{jets} \gtrsim 2 $ and
$ E_T^{miss} > 300 {\rm GeV}$ adopted in \cite{Explanation-2}. This is because we are considering a
compressed spectrum, where the $E_T^{miss}$ of the signal is usually small.
\end{itemize}

\section*{Acknowledgement}
We thank the authors of the CheckMATE, especially Jamie Tattersall, for useful discussions
about the package.
This work was supported by the National Natural Science Foundation of China (NNSFC)
under grant No. 10821504, 11222548, 11121064, 11135003, 90103013 and 11275245,
and by the CAS Center for Excellence in Particle Physics (CCEPP).

\appendix
\section{Validation of Monte Carlo simulations}
\begin{sidewaystable}\caption{Validation table of our implement of the ATLAS search for the leptonic-$Z$ $+ jets + E_T^{miss}$
signal\cite{ATLAS-Z-Excess} in CheckMATE. We considered the process $pp\rightarrow\tilde{g}\tilde{g}\rightarrow\tilde{\chi}_1^0\tilde{\chi}_1^0jjjj\rightarrow\tilde{G}\tilde{G}ZZjjjj$ at 8-TeV LHC
in the GMSB model with two benchmark points taken from \cite{ATLAS-Z-Excess}. Numbers in the columns EXP and OUR are the event numbers of
electron pair or muon pair obtained by the ATLAS collaboration and us respectively, and those in the column DIFF represent their relative difference. \label{tab:valAZ}}
\centering
\footnotesize{
\begin{tabular}{|l||r|r|r||r|r|r||r|r|r||r|r|r|}
\hline
\multicolumn{1}{|c||}{}           & \multicolumn{6}{c||}{$m_{\tilde{g}}=900GeV$, $\mu=600GeV$}                           & \multicolumn{6}{c|}{$m_{\tilde{g}}=1000GeV$, $\mu=700GeV$}                         \\ \hline
                                 & \multicolumn{2}{c|}{EXP}  & \multicolumn{2}{c|}{OUR}   & \multicolumn{2}{c||}{DIFF}  & \multicolumn{2}{c|}{EXP}  & \multicolumn{2}{c|}{OUR}  & \multicolumn{2}{c|}{DIFF}  \\ \hline
No cuts                          & \multicolumn{2}{c|}{189}  & \multicolumn{2}{c|}{189}   & \multicolumn{2}{c||}{}      & \multicolumn{2}{c|}{71.8} & \multicolumn{2}{c|}{71.8} & \multicolumn{2}{c|}{}      \\ \hline
At least 2 leptons               & \multicolumn{2}{c|}{88.8} & \multicolumn{2}{c|}{73.41} & \multicolumn{2}{c||}{-17\%} & \multicolumn{2}{c|}{33.8} & \multicolumn{2}{c|}{26.7} & \multicolumn{2}{c|}{-21\%} \\ \hline
                                 & \multicolumn{3}{c||}{ee}                   & \multicolumn{3}{c||}{$\mu\mu$}           & \multicolumn{3}{c||}{ee}                   & \multicolumn{3}{c|}{$\mu\mu$}          \\ \hline
                                 & EXP         & OUR         & DIFF          & EXP        & OUR           & DIFF       & EXP         & OUR         & DIFF          & EXP       & OUR          & DIFF        \\ \hline
Lepton flavour                   & 36.1        & 30.77       & -15\%         & 25.7       & 28.5          & 11\%       & 14.3        & 11.48       & -20\%          & 9.3       & 10.04        & 8\%        \\ \hline
PromptLeptons                    & 35.3        & 30.73       & -13\%         & 25.6       & 28.43         & 11\%       & 14          & 11.47       & -18\%         & 9.3       & 10.04        & 8\%        \\ \hline
Opposite charged leptons         & 33.6        & 30.05       & -11\%         & 24.2       & 28.01         & 16\%       & 13.3        & 11.27       & -15\%         & 8.8       & 9.94         & 13\%        \\ \hline
$>1$jet                          & 32.2        & 27.75       & -14\%         & 23.1       & 25.97         & 12\%       & 12.8        & 10.48       & -18\%         & 8.4       & 9.07         & 8\%        \\ \hline
$m_{ll}>15$                      & 30.0        & 27.69       & -8\%         & 23.0       & 25.93         & 13\%       & 12.0        & 10.48       & -13\%         & 8.3       & 9.04         & 9\%        \\ \hline
$\Delta\phi(j_1,E_T^{miss})>0.4$ & 28.3        & 25.91       & -8\%         & 21.9       & 24.48         & 12\%       & 11.3        & 9.84        & -13\%         & 8.0       & 8.47         & 6\%        \\ \hline
$\Delta\phi(j_2,E_T^{miss})>0.4$ & 25.7        & 23.27       & -9\%         & 19.9       & 22.13         & 11\%       & 10.3        & 8.82        & -14\%         & 7.2       & 7.66         & 6\%        \\ \hline
$81GeV<m_{ll}<101GeV$            & 22.1        & 21.38       & -3\%         & 16.6       & 18.77         & 13\%       & 8.8         & 8.1         & -8\%         & 5.9       & 6.38         & 8\%        \\ \hline
$H_T>600GeV$                     & 20.5        & 18.01       & -12\%         & 15.1       & 15.54         & 3\%       & 8.4         & 6.94        & -17\%         & 5.6       & 5.52         & -1\%       \\ \hline
$E_T^{miss}>225GeV$              & 15.0        & 13.74       & -8\%         & 11.1       & 11.47         & 3\%       & 6.7         & 5.44        & -19\%         & 4.4       & 4.28         & -3\%       \\ \hline
\end{tabular}
}
\end{sidewaystable}

\begin{table}[thp]\caption{Validation table of our implement of the ATLAS search for the $2\sim6$ jets + $E_T^{miss}$ signal \cite{ATLAS-6jet-new} in CheckMATE.
We considered the process $pp\rightarrow\tilde{g}\tilde{g}\rightarrow\tilde{\chi}_1^\pm\tilde{\chi}_1^\pm jjjj\rightarrow \tilde{\chi}_1^0\tilde{\chi}_1^0 W^\pm W^\pm jjjj$ at 8-TeV LHC
in the MSSM with two benchmark points taken from \cite{ATLAS-6jet-new}. Numbers in the columns EXP and OUR are the cut efficiencies in $2jW$ or $6jt$ signal region obtained by the ATLAS collaboration and us respectively, and those in the column DIFF represent their relative difference. \label{tab:val6jet}}\centering
\begin{tabular}{|l|r|r|r|}
\hline
\multicolumn{4}{|c|}{$\tilde{g}$ $\tilde{g}$ one step, $m_{\tilde{g}} =1200GeV$, $m_{\tilde{\chi}_1^\pm}=1150GeV$, $m_{\tilde{\chi}_1^0}=60GeV $} \\ \hline
             SR:2jW                                            & EXP                     & OUR                     & DIFF                    \\ \hline
$E_T^{miss}>160GeV,~P_T(j_{1,2})>130(60)GeV$                   & 52.70                   & 56.35                   & -7\%                   \\ \hline
$\Delta\phi(j_{1,2,3},E_T^{miss})>0.4$                         & 46.30                   & 49.01                   & -6\%                   \\ \hline
N(W) unresolved $\geq2$                                        & 9.20                    & 8.70                    & 5\%                    \\ \hline
$E_T^{miss}/m_{eff}(N_j)>0.25$                                 & 7.00                    & 6.69                    & 4\%                    \\ \hline
$m_{eff}(incl.)>1800GeV$                                       & 5.30                    & 4.86                    & 8\%                    \\ \hline
\multicolumn{4}{|c|}{$\tilde{g}$ $\tilde{g}$ one step, $m_{\tilde{g}} =1265GeV$, $m_{\tilde{\chi}_1^\pm}=945GeV$, $m_{\tilde{\chi}_1^0}=625GeV$} \\ \hline
 SR:6jt                             & EXP                      & OUR                     & DIFF                    \\ \hline
$E_T^{miss}>160GeV,~P_T(j_1,j_2)>130(60)GeV$                   & 53.30                   & 54.16                   & -2\%                   \\ \hline
$P_T(j_3)>60GeV$                                               & 53.00                   & 53.83                   & -2\%                   \\ \hline
$P_T(j_4)>60GeV$                                               & 50.50                   & 51.50                   & -2\%                   \\ \hline
$P_T(j_5)>60GeV$                                               & 41.40                   & 43.49                   & -5\%                   \\ \hline
$P_T(j_6)>60GeV$                                               & 26.70                   & 29.78                   & -12\%                   \\ \hline
$\Delta\phi(j_{1,2,3},E_T^{miss})>0.4$                         & 22.40                   & 25.38                   & -13\%                   \\ \hline
$\Delta\phi(j_{i>3},E_T^{miss})>0.2$                           & 18.20                   & 20.54                   & -13\%                   \\ \hline
$E_T^{miss}/m_{eff}(N_j)>0.25$                                 & 10.90                   & 11.64                   & -7\%                   \\ \hline
$m_{eff}(incl.)>1500GeV$                                       & 4.20                    & 4.81                    & -15\%                   \\ \hline
\end{tabular}
\end{table}

\begin{table}[thp]\caption{Validation table of our implement of the CMS search for the leptonic-$Z$ $+ jets + E_T^{miss}$ signal
\cite{CMS-Leptonic-Z} in CheckMATE. We considered the process $pp\rightarrow\tilde{g}\tilde{g}\rightarrow\tilde{\chi}_1^0\tilde{\chi}_1^0jjjj\rightarrow\tilde{G}\tilde{G}ZZjjjj$ at 8-TeV LHC
in the GMSB model with the benchmark point taken from \cite{CMS-Leptonic-Z}. Numbers in the columns EXP and OUR are the event numbers of different bins
obtained by the CMS collaboration and us respectively, and those in the column DIFF represent their relative difference.\label{tab:valCZ}}\centering
\begin{tabular}{|l|r|r|r||r|r|r|}
\hline
                                               & \multicolumn{2}{c|}{EXP}  & \multicolumn{2}{c|}{OUR}   & \multicolumn{2}{c|}{DIFF}  \\ \hline
All events                                     & \multicolumn{2}{c|}{37.7} & \multicolumn{2}{c|}{37.7}  & \multicolumn{2}{c|}{0}     \\ \hline
$\geq2$ leptons($l^\pm l^\mp$),$P_{T}>20GeV$   & \multicolumn{2}{c|}{11.9} & \multicolumn{2}{c|}{11.8} & \multicolumn{2}{c|}{-0.5\%} \\ \hline
$81<m_{ll}<101GeV$                             & \multicolumn{2}{c|}{10.7} & \multicolumn{2}{c|}{10.4} & \multicolumn{2}{c|}{-2.7\%}  \\ \hline
                                               & \multicolumn{3}{c||}{$n_{jets}>2$}        & \multicolumn{3}{c|}{$n_{jets}>3$}        \\ \hline
                                               & EXP          & OUR        & DIFF            & EXP         & OUR           & DIFF       \\ \hline
$n_{jets}>2~or~3$                              & 10.7         & 10.4       & -2.9\%          & 10.4        & 10.2          & -1.9\%        \\ \hline
$E_T^{miss}>100GeV$                            & 10.3         & 10.0       & -3.3\%          & 10          & 9.8           & -2.1\%        \\ \hline
$E_T^{miss}>200GeV$                            & 9.2          & 8.8        & -4.1\%          & 8.9         & 8.7           & -2.5\%        \\ \hline
$E_T^{miss}>300GeV$                            & 7.6          & 7.2        & -5.4\%          & 7.4         & 7.1           & -4.3\%        \\ \hline
$100GeV<E_T^{miss}<200GeV$                     & 1.1          & 1.1        & 3\%             & 1.1         & 1.1           & 0.8\%         \\ \hline
$200GeV<E_T^{miss}<300GeV$                     &1.6           & 1.6        & 2\%             & 1.5         & 1.6           & 6.3\%         \\ \hline
\end{tabular}
\end{table}

In this appendix, we show the validations of our simulations on the ATLAS searches for the leptonic-$Z$ $+ jets + E_T^{miss}$ signal \cite{ATLAS-Z-Excess} and the
$2\sim6$ jets + $E_T^{miss}$ signal \cite{ATLAS-6jet-new}. For each search, we choose two benchmark points adopted by the ATLAS report,
and repeat the experimental analysis.
Our simulation results are presented in Table \ref{tab:valAZ} and Table \ref{tab:val6jet} respectively, and all of them indicate that
our simulation agrees with the corresponding ATLAS analysis within $20\%$ uncertainty. We also validated our simulation on the CMS
search for the leptonic-$Z$ $+ jets + E_T^{miss}$ signal
\cite{CMS-Leptonic-Z} with the corresponding results given in Table \ref{tab:valCZ}. This table indicates that we can reproduce the CMS results within $10\%$ uncertainty.

Since the ATLAS preliminary search for $2\sim6$ jets + $E_T^{miss}$ signal \cite{ATLAS-6jet-old} and the search for the $>6$ jets + $E_T^{miss}$ signal \cite{ATLAS-7jet}
do not affect our conclusions, we do not present their validation here. In fact, we checked that we can reproduce these searches
within $15\%$ uncertainties.


\begin{thebibliography}{99}
\bibitem{Higgs-discovery}
  G.~Aad {\it et al.}  [ATLAS Collaboration],
  Phys.\ Lett.\ B {\bf 716}, 1 (2012)
  [arXiv:1207.7214 [hep-ex]];
   S.~Chatrchyan {\it et al.}  [CMS Collaboration],
  Phys.\ Lett.\ B {\bf 716}, 30 (2012)
  [arXiv:1207.7235 [hep-ex]].

\bibitem{ATLAS-Z-Excess}
  G.~Aad {\it et al.}  [ATLAS Collaboration],
  arXiv:1503.03290 [hep-ex].

\bibitem{Explanation-1}
  G.~Barenboim, J.~Bernabeu, V.~A.~Mitsou, E.~Romero, E.~Torro and O.~Vives,
  arXiv:1503.04184 [hep-ph].

\bibitem{Explanation-2}
  U.~Ellwanger,
  arXiv:1504.02244 [hep-ph].

\bibitem{Explanation-3}
  B.~Allanach, A.~Raklev and A.~Kvellestad,
  arXiv:1504.02752 [hep-ph].

\bibitem{Explanation-4}
  A.~Kobakhidze, A.~Saavedra, L.~Wu and J.~M.~Yang,
  arXiv:1504.04390 [hep-ph].

\bibitem{Explanation-5}
  We note that the letponic-$Z$ excess may also be explained
  in the context of composite Higgs / RS theories, see for example
    N.~Vignaroli,
  arXiv:1504.01768 [hep-ph].

\bibitem{GGM}
  P.~Meade, N.~Seiberg and D.~Shih,
  Prog.\ Theor.\ Phys.\ Suppl.\  {\bf 177}, 143 (2009)
  [arXiv:0801.3278 [hep-ph]].

\bibitem{GGM-Light-G}
   J.~L.~Feng, S.~Su and F.~Takayama,
  Phys.\ Rev.\ D {\bf 70}, 075019 (2004)
  [hep-ph/0404231].

\bibitem{NMSSM}
  U.~Ellwanger, C.~Hugonie and A.~M.~Teixeira,
  Phys.\ Rept.\  {\bf 496}, 1 (2010)
  [arXiv:0910.1785 [hep-ph]].

\bibitem{NMSSMTools}
  U.~Ellwanger, J.~F.~Gunion and C.~Hugonie,
  JHEP {\bf 0502}, 066 (2005)
  [hep-ph/0406215].

\bibitem{Higgs-Mass-1}
  U.~Ellwanger,
  JHEP {\bf 1203}, 044 (2012)
  [arXiv:1112.3548 [hep-ph]].

\bibitem{Higgs-Mass-2}
  J.~ Cao {\it et al.},
  JHEP {\bf 1203}, 086 (2012)
  [arXiv:1202.5821 [hep-ph]].

\bibitem{Higgs-Mass-3}
  S.~F.~King, M.~Muhlleitner and R.~Nevzorov,
  Nucl.\ Phys.\ B {\bf 860}, 207 (2012)
  [arXiv:1201.2671 [hep-ph]].

\bibitem{Higgs-Mass-4}
  J.~F.~Gunion, Y.~Jiang and S.~Kraml,
  Phys.\ Lett.\ B {\bf 710}, 454 (2012)
  [arXiv:1201.0982 [hep-ph]].

\bibitem{NMSSM-Dark-Matter}
  J.~Cao {\it et al.},
  JHEP {\bf 1405}, 056 (2014)  [arXiv:1311.0678 [hep-ph]].

\bibitem{ATLAS-6jet-old}
  The ATLAS collaboration,
  ATLAS-CONF-2013-047, ATLAS-COM-CONF-2013-049.

\bibitem{ATLAS-6jet-new}
  G.~Aad {\it et al.}  [ATLAS Collaboration],
  JHEP {\bf 1409}, 176 (2014)
  [arXiv:1405.7875 [hep-ex]].

\bibitem{ATLAS-7jet}
  G.~Aad {\it et al.}  [ATLAS Collaboration],
  JHEP {\bf 1310}, 130 (2013)
  [arXiv:1308.1841 [hep-ex]].

\bibitem{CMS-Leptonic-Z}
  V.~Khachatryan {\it et al.}  [CMS Collaboration],
  arXiv:1502.06031 [hep-ex].

\bibitem{antikt}
  M.~Cacciari, G.~P.~Salam and G.~Soyez,
  JHEP {\bf 0804}, 063 (2008)
  [arXiv:0802.1189 [hep-ph]].

\bibitem{CLs}
  A.~L.~Read,
  J.\ Phys.\ G {\bf 28}, 2693 (2002).

\bibitem{checkmate}
  M.~Drees {\it et al.},
  Comput.\ Phys.\ Commun.\  {\bf 187}, 227 (2014)
  [arXiv:1312.2591 [hep-ph]].
   J.~S.~Kim, D.~Schmeier, J.~Tattersall and K.~Rolbiecki,
  arXiv:1503.01123 [hep-ph].

\bibitem{delphes}
  J.~de Favereau {\it et al.}  [DELPHES 3 Collaboration],
  JHEP {\bf 1402}, 057 (2014)
  [arXiv:1307.6346 [hep-ex]].

\bibitem{mg5}
  J.~Alwall {\it et al.},
  JHEP {\bf 1106}, 128 (2011)
  [arXiv:1106.0522 [hep-ph]].

\bibitem{pythia}
  T.~Sjostrand, S.~Mrenna and P.~Z.~Skands,
  JHEP {\bf 0605}, 026 (2006)
  [hep-ph/0603175].

\bibitem{prospino}
  W.~Beenakker, R.~Hopker and M.~Spira,
  hep-ph/9611232.

\end{thebibliography}
\end{document}